\documentclass[12pt]{article}
\usepackage{pic03}
\usepackage{hyperref}
\usepackage{url}
\usepackage{graphicx}

\begin{document}

\title{\bf
\textit{FINAL RESULTS FROM DELPHI ON THE SEARCHES FOR SM AND MSSM
NEUTRAL HIGGS BOSONS}}
\author{
Javier Fern{\'a}ndez\\
{\em Instituto de F{\'i}sica de Cantabria, Univ. de Cantabria} \\
On behalf of the DELPHI Collaboration}
\maketitle

\baselineskip=14.5pt
\begin{abstract}
These final results from DELPHI searches for the Standard Model SM
Higgs boson, together with benchmark scans of the Minimal
Supersymmetric Standard Model MSSM neutral Higgs bosons, used data
taken at centre-of-mass energies between 200 and 209 GeV with a
total integrated luminosity of 224 $pb^-1$.The data from 192 to
202 GeV are reanalyzed with improved b-tagging for MSSM final
states decaying to four b-quarks. The 95\% confidence level lower
mass bound on the Standard Model Higgs boson is 114.1 GeV. Mass
limits are also given on the lightest scalar and pseudo-scalar
Higgs bosons of the MSSM.

\end{abstract}

\baselineskip=17pt

\section{Introduction}

The dominant production mechanism at LEP for a scalar Higgs boson,
such as the {\sc SM} predicts, is  the  s-channel process
${\mathrm e}^+{\mathrm e}^- \rightarrow \mbox{Z}^* \rightarrow
HZ$. In the {\sc MSSM}, the production of the lightest scalar
Higgs boson, h, proceeds through the same processes as in the {\sc
SM}.

The data from the search for the {\sc SM} Higgs boson also provide
information on  the h boson. However, in the {\sc MSSM} the
production cross-section is  smaller than the {\sc SM} one and can
even vanish in certain  regions of the {\sc MSSM} parameter space.
There is also a CP-odd pseudo-scalar, A, which would be produced
mostly in the $e^+e^- \rightarrow \mbox{Z}^* \rightarrow {\mathrm
h} {\mathrm A}$ process at LEP2. This channel is therefore also
considered. For {\sc MSSM} parameter values for which single h
production is suppressed, the associated $hA$ production is
enhanced (if kinematically permitted).

 In the $HZ$ channel, all known decays of the Z boson (hadrons,
charged leptons and neutrinos) have been taken into account, while
the analyses have been optimized for decays of the Higgs particle
into \mbox{${\mathrm b}\bar{\mathrm b}$}, making use of the
expected high branching fraction of this mode, and for Higgs boson
decays into a pair of \mbox{$\tau$} particles, which is the second
main decay channel in the  {\sc SM} and in most of the {\sc MSSM}
parameter space. The $hA$ production has been searched for in the
two main decay channels, namely the $b\bar{b} ~ b \bar{b}$  and
\mbox{${\mathrm b}\bar{\mathrm b}\tau^+ \tau^-$} final states.

The present results are published in~\cite{DELPHI}.The {\sc SM}
Higgs limit is improved when combining the four LEP experiments to
$m_H > 114.4 GeV/c^2$~\cite{LEP}. A LEP combination in the {\sc
MSSM} is expected for the coming months.

\section{Statistical procedure}
The statistical procedure to combine the different search channels
is a confidence level calculation in two hypothesis: a background
only and a background plus signal hypothesis. Confidence levels
are calculated using a modified frequentist technique based on the
extended maximum likelihood ratio~\cite{alex} which has also been
adopted by the LEP Higgs working group.

The basis of the calculation is the likelihood ratio
test-statistic defined using the signal and background densities
for each event. These densities are constructed using
two-dimensional discriminant information. The first variable is
the reconstructed Higgs boson mass (or the sum of the
reconstructed h and A masses in the $hA$ channels), the second one
is channel-dependent.

\section{SM Higgs}
\begin{figure}[htbp]
  \centerline{\hbox{ \hspace{0.2cm}
    \includegraphics[width=8cm]{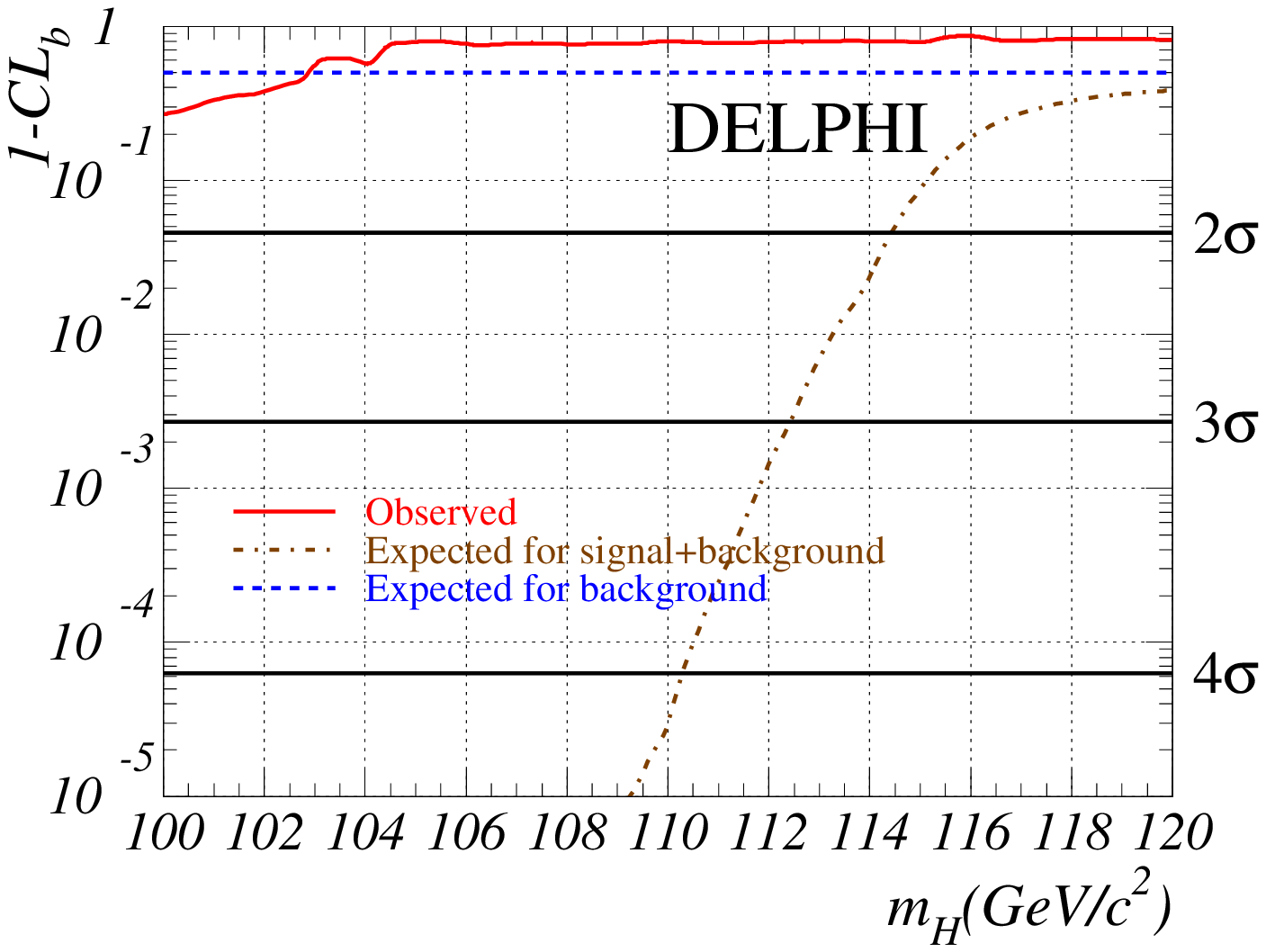}
    \includegraphics[width=8cm]{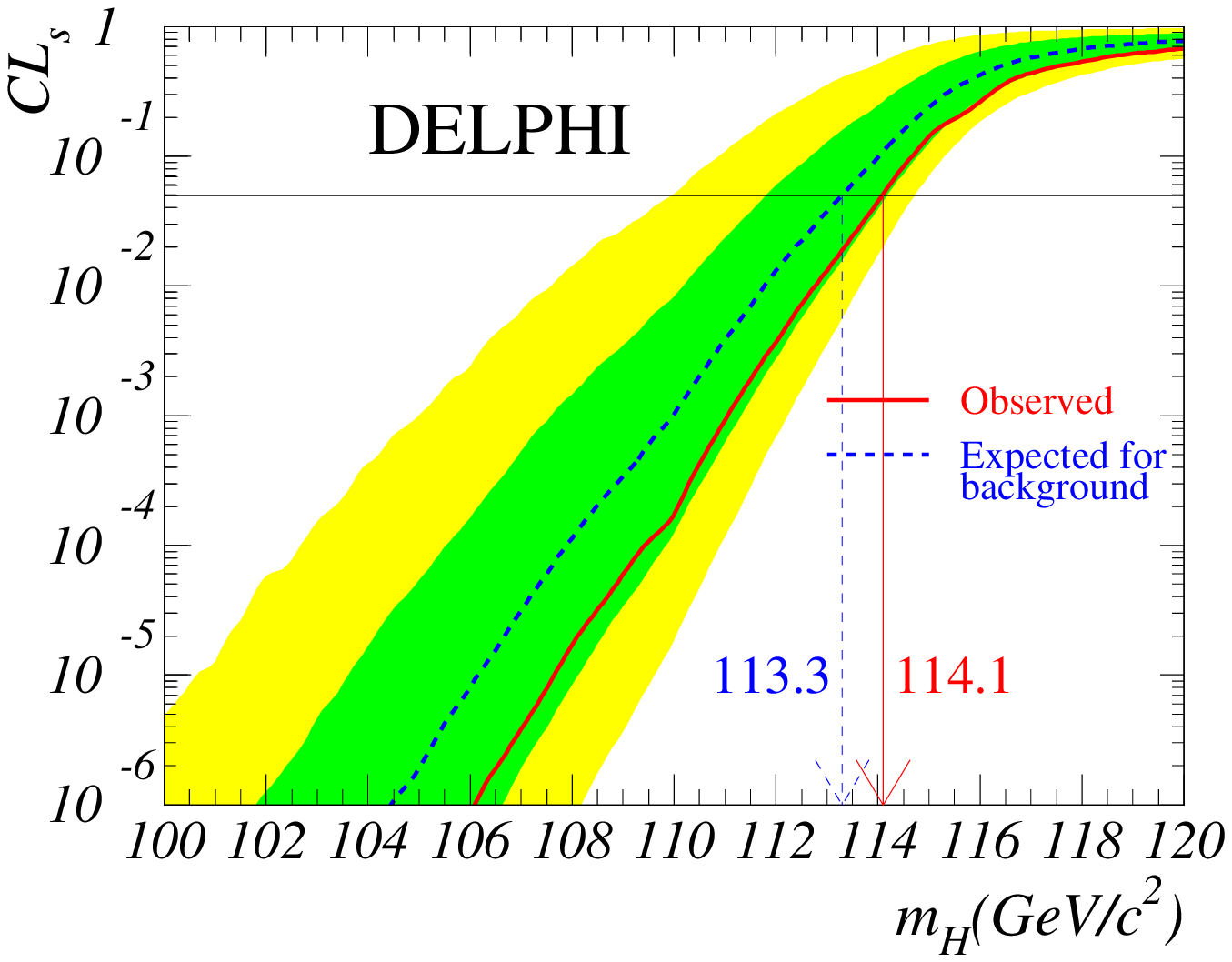}
    }
  }
 \caption{\it
      Final DELPHI {\sc SM} Higgs $CL_b$ and $CL_s$ curves.
    \label{sm} }
\end{figure}

Curves of the confidence level $CL_b$ and $CL_s$  as a function of
the test mass $m_H$ are shown in Fig.~\ref{sm}. In the presence of
a sizable Higgs signal, the value of the observed $CL_b$ would
approach one, since it measures the fraction of experiments with
only background processes which are more background-like than the
observation. Here the compatibility between the observation and
the expectation from background processes is well within one
standard deviation over the range of masses tested. The observed
95\% {\sc CL} lower limit on the mass is 114.1~$GeV/c^2$ while the
expected median limit is $113.3~GeV/c^2$.

\section{MSSM neutral Higgs}
In the benchmark scenarios proposed, a scan was performed over the
{\sc MSSM} parameters $tan \beta$ and $m_A$. The results come in
terms of exclusion ranges as shown in Table~\ref{extab}. An
updated version of the {\sc MSSM} DELPHI results can be found
in~\cite{vanina}.

\begin{table}[htbp]
\centering \caption{ \it Final DELPHI {\sc MSSM} Neutral Higgs
exclusion ranges. } \vskip 0.1 in
\begin{tabular}{|c|c|} \hline

          \multicolumn{2}{|c|}{$m_h^{max}$ scenario:} \\
\hline
 $m_h > 89.7 GeV/c^2$ and $m_A > 90.5 GeV/c^2$ & for any $tan \beta \geq 0.4$\\
$tan \beta < 0.54$ or $tan \beta > 2.36$ & for any $m_A$\\
\hline
          \multicolumn{2}{|c|}{no mixing scenario scenario:} \\
 \hline
$m_h > 92.0 GeV/c^2$ and $m_A > 93.0 GeV/c^2$ & for any $tan \beta \geq 0.8$\\
$tan \beta < 0.8$ or $tan \beta > 9.36$ & for any $m_A$\\
\hline
\end{tabular}
\label{extab}
\end{table}

\end{document}